# Enhancing Power Flow Estimation with Topology-Aware Gated Graph Neural Networks

Shrenik Jadhav[1], Birva Sevak[1], Srijita Das[1], Wencong Su[2, *], and Van-Hai Bui[2, *]

[1]Departmernt of Computer and Information Science, University of Michigan-Dearborn

[2]Departmernt of Electrical and Computer Engineering, University of Michigan-Dearborn

**Abstract:** Accurate and scalable surrogate models for AC power flow are essential for real-time grid monitoring, contingency analysis, and decision support in increasingly dynamic and inverter-dominated power systems. However, most existing surrogates fall short of practical deployment due to their limited capacity to capture long-range nonlinear dependencies in meshed transmission networks and their weak enforcement of physical laws. These models often require extensive hyperparameter tuning, exhibit poor generalization under topology changes or large load swings, and typically do not quantify uncertainty or scale well beyond a few hundred buses. To address these challenges, this paper proposes a gated graph neural network (GGNN) surrogate for AC power-flow estimation under topological uncertainty. The model is trained across multiple IEEE benchmark networks of varying size and complexity, each incorporating randomized line contingencies and up to 40% load variation. To improve robustness and generalization, we explore both conventional supervised learning and physics-informed self-supervised training strategies. Comparative evaluations show that the proposed GGNN consistently outperforms prior GNN-based surrogates, achieving predictions closely aligned with Newton–Raphson solutions. By embedding operational constraints directly into the architecture and loss function, the model ensures physical consistency and delivers a lightweight, accurate, and scalable tool for real-time grid operations.

## 1.Introduction

Accurate power-flow analysis is foundational for operating and planning transmission networks. The AC power-flow equations, which govern bus voltages and branch flows, are traditionally solved using iterative numerical methods such as the Newton–Raphson algorithm and, later, the fast-decoupled formulation [1], [2]. For instance, the Midcontinent Independent System Operator (MISO) reports that screening every N-1 contingency on its 20,000-bus footprint involves performing Real-Time Contingency Analysis on over 10,000 contingency cases every five minutes, which is equivalent to more than 120,000 AC power-flow solves per hour, and imposes a real-time computational load of approximately 20 CPU-hours per hour of operation even with an optimized Newton–Raphson implementation [3]. As renewable energy integration increases variability and uncertainty in power system behavior, the number of probabilistic scenarios to evaluate grows dramatically, resulting in a cumulative computational burden that can quickly become intractable for large-scale systems [4]. Consequently, there has been growing interest in developing surrogate models that can approximate AC power-flow solutions with significantly reduced computational cost while maintaining high accuracy.

Recent years have therefore seen a surge of data-driven and graph-based methods to accelerate power-flow computation. While simplified DC power-flow approximations offer 5–10× computational speed-ups, they do so at the expense of reactive power fidelity. In contrast, emerging machine learning techniques promise near-AC accuracy without requiring iterative solvers. Among these, graph neural networks (GNNs) are particularly attractive because they exploit the inherent graph structure of power grids. Early work by Owerko et al. [5] proved that GNNs can learn optimal power flow mappings; Yang et al. recently advanced the message-passing paradigm by introducing a physics-guided GNN that performs full and even probabilistic AC power-flow estimation with Monte-Carlo–level fidelity [6], and Nakiganda & Chatzivasileiadis showed its utility for fast N-1 screening [7]. The 2023–2025 literature reflects rapid advancements in surrogate modeling using increasingly sophisticated GNN architectures. For instance, Lin et al. [8] introduced PowerFlowNet, a physics-informed message-passing GNN that matches Newton–Raphson accuracy while running four times faster on the IEEE-14 bus system and 145 times faster on a 6470-bus French grid; Eeckhout et al. incorporate a physics-aware loss to improve out of sample robustness on a 3-bus feeder [9]; PowerGNN's topology recurrent architecture [10] reaches 0.13–0.17 per unit RMSE on the NREL-118 system (converted from its reported 2.2–2.8 kV RMS on a 13 kV base), yet still misses tight reactive-power limits; Talebi & Zhou [11] compare four backbone designs and report approximately 25× CPU speed-ups over Newton–Raphson; Zhou et al. embedded Kirchhoff constraints into a graphical state-estimator but observe increasing angle errors on IEEE-300 [12]. Chen et al. [13] introduced Powerformer, a transformer-based surrogate that reaches GNN-level accuracy on large power-grid cases but needs significantly more GPU memory than lightweight GNNs; Authier et al. [14] embed power-flow physics directly into *GraPhyR*, a GNN for real-time distribution-grid reconfiguration, demonstrating feasible solutions on multiple feeder topologies up to 34 buses; Ghamizi et al. [15] introduced *SafePowerGraph*, a safety-oriented benchmark that measures physical-constraint violations and prediction error across several GNN architectures for transmission-grid contingency analysis, while Wu and Xu [16] applied transfer learning in their TGACN-TL framework to generalize multi-energy-flow calculations across integrated electricity, gas, and heating networks, demonstrating robust accuracy under changing topologies and uncertainties.

Despite recent advancements, several critical gaps remain in the development and evaluation of power-flow surrogates. As highlighted in a recent survey by Liao et al. [17] and the OPF Data benchmark suite [18], most published results are limited to systems no larger than IEEE-300, leaving large-scale scalability largely unexplored. Early fully connected architectures, such as the 20-layer feed-forward network proposed by Baranwal et al. [19], required complete retraining when the topology changed, significantly hindering real-world applicability. Robustness under extreme operating conditions also remains a concern. For instance, Gao et al. [20] tested their physics-guided GCN with bus loads varied by ±20 % and kept voltage-magnitude errors below 1 %. Yet most existing surrogates still give only one deterministic prediction, with no estimate of uncertainty; Bayesian-GNN work by Rivera-Ortega [21] provides uncertainty intervals but is restricted to a 57-bus feeder. Purely data-driven predictors can additionally violate operational limits, as surveyed by Mohammadi et al. [22]. Even with graph-aware training strategies designed to generalize across multiple network configurations, challenges persist. Hansen et al. [23], for example, report mean absolute voltage errors exceeding 2% on IEEE-300 using a decentralized line-graph GNN.

In summary, most existing surrogate models fall short of practical deployment. Relatively shallow or purely feed forward architecture often fail to capture the long-range nonlinear dependencies of meshed transmission grids. Due to weak enforcement of physical constraints, these models frequently violate constraints during N-1 contingencies or large load swings. Additionally, their training pipelines demand extensive hyper-parameter searches and hundreds of epochs, hindering rapid retraining as grid states evolve. Uncertainty quantification is rarely addressed, and scalability beyond a few hundred buses remains elusive.

Therefore, in this paper, a gated graph neural network (GGNN) surrogate is proposed for alternating current power flow estimation that maintains high accuracy under significant topological uncertainty. The model is trained on multiple IEEE benchmark networks of varying size and complexity, converging in approximately 7 hours on standard CPU for our largest bus case, with each scenario featuring randomized line contingencies and up to approximately 40% of load-demand variations. To improve model robustness and generalization, we evaluate a range of training strategies, including conventional supervised learning and physics-informed self-supervised approaches. In comparative experiments, the proposed GGNN consistently outperforms prior GNN-based surrogates, delivering AC power-flow predictions that closely match the true Newton–Raphson solution. By embedding operational constraints directly into the network design and loss function, the surrogate model inherently satisfies the underlying power-flow physics. The result is a lightweight, physics-consistent AC power-flow solver that delivers solutions with negligible loss of accuracy, making it ideally suited for real-time contingency analysis and grid management.

## 2. System Models

### 2.1 Graph-Based Reformulation of Power Flow Problem

The conventional AC power flow (PF) problem is defined over a power transmission network, where the objective is to compute the steady state voltage magnitudes $V_i$ and voltage phase angles $\theta_i$ at each bus $i \in \{1, \dots, N\}$. These variables are governed by non-linear algebraic equations derived from Kirchhoff's laws and the power balance constraints [24]:

$$P_i = \sum_{j=1}^{N} |V_i||V_j|(G_{ij} \cos\theta_{ij} + B_{ij} \sin\theta_{ij}) \tag{1}$$

$$Q_i = \sum_{j=1}^{N} |V_i||V_j|(G_{ij} \sin\theta_{ij} - B_{ij} \cos\theta_{ij}) \tag{2}$$

where $P_i$ and $Q_i$ are the net active and reactive power injections at bus $i$, $G_{ij}$ and $B_{ij}$ are the real and imaginary parts of the bus admittance matrix Ybus, and $\theta_{ij} = \theta_i - \theta_j$. Classical Newton–Raphson solvers iteratively linearize these equations, but their performance can degrade under frequent topology changes or real-time requirements.

The current and complex power injected into buses are expressed in (3) and (4) respectively.

$$I = Y_{bus}V \tag{3}$$

$$S_i = P_i + jQ_i = V_i I_i = V_i (\Sigma_j Y_{ij} V_j) \tag{4}$$

Linearizing the mismatch $\Delta S = [\Delta P; \Delta Q]$ around the current iterate yields the NR update

$$\Delta S = [J][\Delta\theta; \Delta|V|] \tag{5}$$

where J is the four-block Jacobian of partial derivatives $(\partial P/\partial\theta, \partial P/\partial|V|, \partial Q/\partial\theta, \partial Q/\partial|V|)$. Repeated solution of (5) dominates run-time on large systems.

To enable scalable and adaptive learning-based alternatives, we reformulate the power flow problem on a graph $G = (V, E)$, Each node $v_i \in V$ represents a bus and each edge $e_{ij} \in E$ represents the line or transformer between buses $i$ and $j$. This topology captures both local Kirchhoff couplings and global mesh connectivity, making it well suited for GNN. A GNN surrogate learns the mapping from node- and edge-level electrical features to the bus voltages, thereby replacing the costly iterative solution of (5) with a single forward pass on the graph.

## 2.2 Node and Edge Feature Design

Each node $v_i$ is initialized with a feature vector $x_i \in \mathbb{R}^d$ constructed from physically meaningful attributes:

$$x_i = [P_i, Q_i, V_i, \theta_i, PQ_i, PV_i, Slack_i] \tag{6}$$

These features include power injections, initial voltage estimates (which may be flat start or derived from another solver), and categorical indicators of bus type. Optionally, edge features $e_{ij}$ may encode line parameters like admittance magnitude or phase shift. Figure 1 illustrates the graph-based node–edge abstraction and the conventional single-line schematic of the IEEE 30-bus system, highlighting how the described node and edge features are assigned.

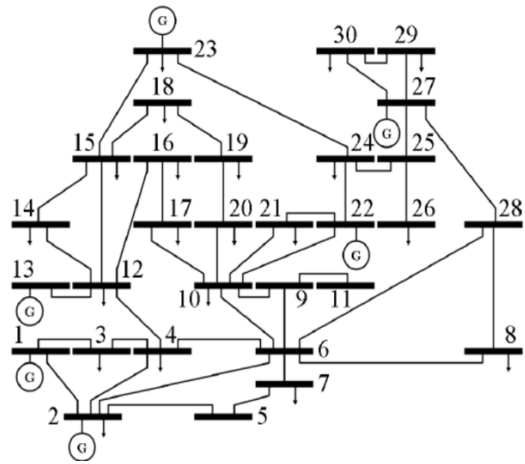

(a)

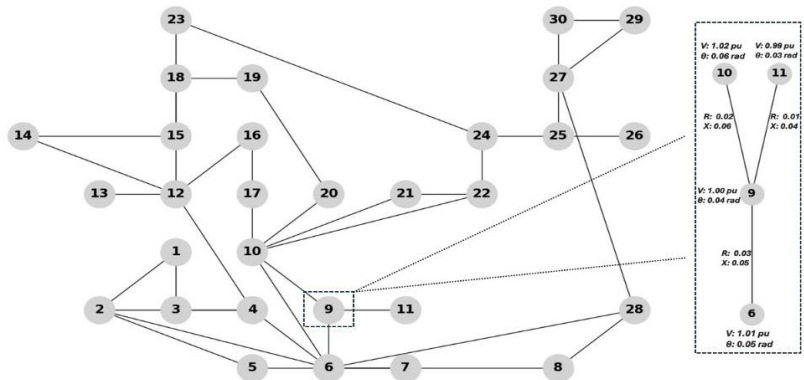

(b)

**Figure 1:** *(a) Electrical single-line diagram preserves the physical topology used for traditional simulation and control [25]. (b) Graph-based representation abstracts buses and lines into nodes and edges, enabling GNN-based message passing.*

The output of the GNN is a learned function $f_\theta$ that maps the entire graph representation to the desired bus-level predictions:

$$f_\theta : \{x_i, e_{ij}\}_{i,j} \rightarrow \{\widehat{V}_i, \widehat{\theta}_i\}_i \tag{7}$$

This task is structured as a supervised regression problem where the model is trained on historical or simulation-based datasets of solved power flow states. The loss function typically measures mean squared error (MSE) between predicted and true values [26]:

$$L = \frac{1}{N} \sum_{i=1}^{N} \left[ (\widehat{V}_i - V_i)^2 + (\widehat{\theta}_i - \theta_i)^2 \right] \tag{8}$$

## 2.3 Proposed Gated GNN-based Power Flow Estimation

### 2.3.1. Conventional Graph Neural Networks

GNNs extend neural networks to graph-structured data by using message passing between connected nodes. In a conventional GNN layer, each node receives information from its neighbors and updates its state through an aggregation and transformation process. Formally, for a node *i* at layer *l*, a generic update can be written as [27]:

$$h_i^{(l+1)} = \sigma \left( W^{(0)} h_i^{(l)} + \sum_{j \in N(i)} W^{(1)} h_j^{(l)} \right) \tag{9}$$

where $h_i^{(l)}$ is the feature vector of node *i* at layer *l*, $N(i)$ is the set of neighbors of *i*, and $W^{(0)}$ and $W^{(1)}$ are learnable weight matrices for the node itself and its neighbors, respectively.

The function $\sigma(\cdot)$ denotes nonlinear activation. This formulation means each node's new embedding $h_i^{(l+1)}$ is computed from a weighted combination of its own previous embedding and the embeddings of its adjacent nodes. Stacking multiple such GNN layers allows information to propagate through the graph, each layer expanding the receptive field by one hop. Conventional GNNs typically perform a fixed number of message-passing layers with no recurrent internal state. Each layer has its own weights, and the node features are updated in one pass per layer.

### 2.3.2. Gated Graph Neural Networks (GGNNs)

In this section, we propose using GGNN with recurrent mechanisms to overcome some limitations of conventional GNNs. In early formulations of GNNs, one needed to iterate until convergence or imposing constraints (like contraction mapping) to guarantee stable node representations. GGNNs, proposed by Li et al. [28], avoid these restrictions by using a gated recurrent unit (GRU) to iteratively update the node states over a fixed number of time steps $T$, like unrolling a recurrent neural network on the graph. Instead of a fixed two- or three-layer feedforward propagation, a GGNN performs iterative message passing with shared weights and gating, which allows the model to refine node embeddings gradually and handle long-range dependencies more effectively.

In GGNN, each node $i$ maintains a hidden state $h_i^{(t)}$ at iteration $t$. Initially, $h_i^{(0)}$ is set based on the node's input features (for example, one can initialize $h_i^{(0)}$ as a learned projection of the feature vector, or simply the feature vector padded to the hidden state size). At each time step $t = 1, 2, \ldots, T$, every node receives messages from its neighbors and updates its hidden state through a GRU. Message aggregation can be written as:

$$m_i^{(t)} = \sum_{j \in N(i)} W_m h_j^{(t-1)} \tag{10}$$

where $W_m$ is a weight matrix for incoming neighbor messages (shared across iterations). Then the new hidden state is produced by the GRU taking the previous state and the aggregated message:

$$h_i^{(t)} = GRU\left(m_i^{(t)}, h_i^{(t-1)}\right) \tag{11}$$

where $GRU(\cdot)$ denotes the gating operations applied to combine the prior state and new information. Internally, the GRU computes update and reset gates to determine how much of the past state to retain and how much new message information to incorporate. Specifically, for each node $i$ at time $t$:

$$\text{Update gate: } z_i^{(t)} = \sigma\left(W_z m_i^{(t)} + U_z h_i^{(t-1)}\right) \tag{12}$$

$$\text{Reset gate: } r_i^{(t)} = \sigma\left(W_r m_i^{(t)} + U_r h_i^{(t-1)}\right) \tag{13}$$

$$\text{Candidate state: } \hat{h}_i^{(t)} = \tanh\left(W_h m_i^{(t)} + U_h\left(r_i^{(t)} \odot h_i^{(t-1)}\right)\right) \tag{14}$$

$$\text{New state: } h_i^{(t)} = z_i^{(t)} \odot h_i^{(t-1)} + \left(1 - z_i^{(t)}\right) \odot \hat{h}_i^{(t)} \tag{15}$$

where, $W_z, U_z, W_r, U_r, W_h, U_h$ are learnable weight matrices, $\sigma$ is the sigmoid function, and $\odot$ denotes element-wise multiplication. These gating equations ensure that each node can decide to keep part of its previous state and update only the necessary components based on new messages. The iteration is repeated for T steps, reusing the same weights at each step; in other words, it unrolls a single layer T time, and gradients are computed through time via backpropagation through time. This gated recurrent architecture enables information to propagate across up to T hop neighborhoods while mitigating the vanishing and exploding gradient problems that arise when stacking many plain GNN layers.

In summary, the key differences are that conventional GNNs use a feed forward propagation with a fixed number of layers and typically no gating, whereas GGNNs use recurrent message passing with gating to iteratively refine node states. GGNNs effectively incorporate an LSTM/GRU like memory into each node, making them well-suited to capturing complex relationships and even sequential outputs on graphs. In the context of our problem, we explore both approaches: a standard multi-layer GNN model and a GGNN model with gated message passing, as described next.

### 2.3.3. GGNN-based Power Flow Estimation with Stability Focused Training

For the task of predicting bus voltages in a power grid, we first represent the electrical network as a graph $G = (V, E)$ where each bus (node) $i \in \mathcal{V}$ corresponds to a node in the graph and each transmission line or transformer between buses corresponds to an edge $(i, j) \in E$ connecting the respective nodes. This graph structure allows the GNN to naturally incorporate the connectivity of the power grid (which buses are directly connected) into the learning process. Each bus (node) is associated with a feature vector capturing its electrical properties and known inputs. In our formulation, the node feature vector for bus $i$ includes active power injection $P_i$, reactive power injection $Q_i$, the initial or specified voltage magnitude $v_i$, the initial voltage phase angle $\theta_i$, and indicator variables for the bus type ($Slack, PV, or\ PQ\ type$). These features summarize the local state and parameters of each bus. For example, the slack bus (reference bus) has a fixed voltage magnitude and angle (serving as the reference), $PV$ (generator) buses have fixed voltage magnitude and specified power, and $PQ$ (load) buses have specified $P$ and $Q$. The inclusion of bus-type indicators (one-hot flags for $Slack/PV/PQ$) helps the model understand which inputs are held constant or controlled for that bus [29]. All features are normalized to appropriate ranges before inputting them into the network.

We begin with a conventional graph-convolutional baseline that stacks two first-order GCN layers followed by a global multi-layer perceptron (MLP). After two such layers (dimension 30 × 12), the node embeddings are flattened (30 × 12 → 360) and then passed through two dense layers (360 → 128 → 60) to yield voltage magnitude and angle predictions $[V_1, \theta_1, V_2, \theta_2, \ldots, V_{30}, \theta_{30}]$. Figure 2 illustrates this pipeline. Although effective, this single-pass architecture cannot explicitly refine its estimates and may struggle with complex non-linearities present in large grids.

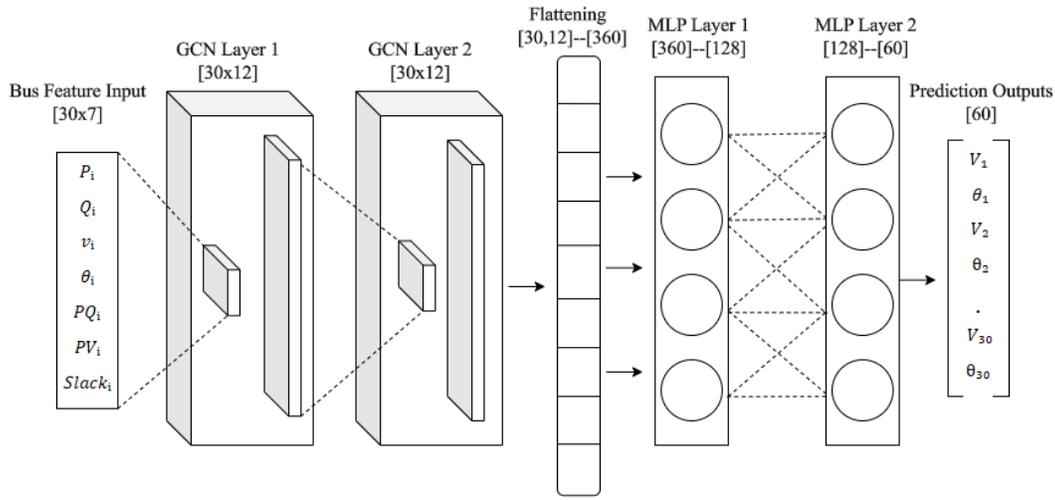

**Figure 2:** *Graph-based model for IEEE 30-bus voltage prediction: inputs pass through two GNN layers and MLP to estimate voltage magnitude and angle at each bus.*

To address these limitations, we adopt a GGNN that performs iterative, gate-controlled message passing analogous to the numerical refinement steps of traditional power-flow solvers. The full GGNN formulation, training strategy, and stability-oriented hyper-parameters are detailed below.

We adopted a GGNN architecture to model the AC power flow in the grid, treating buses as nodes and transmission lines as edges. This existing GGNN model employs a gated recurrent unit GRU for iterative message passing. In each propagation step illustrated in Figure 3, every node bus aggregate "messages" from its neighboring nodes (e.g. combining the hidden states of adjacent buses) and then updates its own hidden state via a GRU cell. The GRU's gating mechanism regulates how much new neighbor information is incorporated versus retained from the previous state, which stabilizes the iterative update. The initial hidden state for each node is obtained by projecting the node's input features, such as bus type and initial electrical quantities into the latent space. A fixed number of these message-passing iterations (i.e. unrolled "time" steps) are executed, rather than iterating to convergence making the process a deterministic RNN-like propagation that is amenable to backpropagation. This setup mirrors the physical power flow interactions, as each bus incrementally assimilates neighbor information, analogous to how electrical quantities propagate in a network.

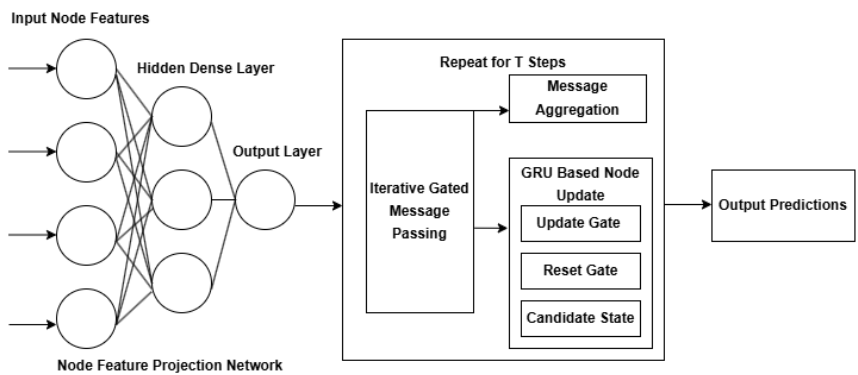

**Figure 3:** *Gated message-passing pipeline in a GGNN: input node features are linearly projected, then refined over T iterative rounds of message aggregation and GRU-based updates (update/reset gates) to produce the final prediction.*

After the iterative message passing, a readout layer produces the voltage predictions for each bus. Specifically, the final hidden state of each node is passed through a feed-forward output model to predict that node's voltage magnitude $v_i$ and phase angle $\theta_i$. In this way, the GGNN effectively learns to approximate the nonlinear AC power flow mapping from inputs to the power flow solution (bus voltage magnitudes and angles). The use of a GRU-based graph model is advantageous here because it can naturally encode the power grid's topology and enforce an iterative refinement of node voltages, much like the successive updates in traditional power flow algorithms. Notably, we did not devise new architecture for this task rather, we leveraged the proven GGNN framework from prior literature and tailored it to the power flow context. As shown in Figure 3 the model's gated message-passing scheme allows information to propagate through many hops in the network while mitigating unstable updates, which is crucial for capturing the complex, non-linear interdependencies of AC power flow.

Training this GGNN model required a stability focused configuration to ensure robust learning. We used an exceedingly small learning rate $5 \times 10^{-5}$ (with the Adam optimizer) [30], markedly lower than the $10^{-3}$–$10^{-2}$ range common in GNN training to slow down learning and avoid divergence on the nonconvex loss surface. A moderate dropout regularization of 10–20% was applied to the neural layers to prevent overfitting and improve robustness [31]. We also applied an L2 weight decay of $1 \times 10^{-6}$ on the parameters, further constraining the model capacity. A batch size of 16 is chosen to balance gradient noise and convergence quality. Additionally, gradient clipping was employed during backpropagation to cap any excessively large gradient norms, thereby preventing exploding updates in the GRU parameters [32]. These cautious settings were experimentally chosen to cope with the highly nonlinear relationships in AC power flow and to promote stable convergence of the model.

Our training regimen deliberately diverges from typical GNN practices in favor of improved generalization and stability. For example, conventional graph neural networks often train with higher learning rates and substantial dropout 50% and weight decay 5e-4 as defaults [33], but such aggressive settings led to erratic performance in our initial trials on AC power flow. By contrast, our conservative setup including early stopping with a patience of 100 epochs and a *ReduceLROnPlateau* scheduler to automatically reduce the learning rate upon plateauing validation loss, yielded a more reliable training process [34]. The model exhibited smooth learning curves and avoided overfitting, achieving better generalization to unseen grid scenarios. In particular, the combination of GRU-based message passing and cautious training hyperparameters produced a network that remains stable when predicting voltages under varying operating conditions. This stability focused training approach ultimately improves the GGNN's ability to capture the complex nonlinear power flow patterns without sacrificing accuracy, resulting in more trustworthy voltage magnitude and angle predictions across a wide range of grid states.

## 3. Experiments and Results

### 3.1 IEEE Test Systems

To evaluate the effectiveness and scalability of our proposed GGNN-based power flow solver, we employ four IEEE benchmark transmission systems: the 30-bus, 118-bus, 300-bus, and 1354-bus networks. These test cases are well-established in power systems literature and are frequently used for validating load flow algorithms under steady-state operating conditions [35], [36]. Each system introduces increasing complexity in terms of bus count, interconnections, and power injection patterns, allowing us to systematically assess the model's performance across small, medium, and large-scale grids. Table 1 summarizes their key characteristics.

**Table 1:** *Summary of IEEE Test Systems Used for Power Flow Analysis*

| System | Buses | Branches | Generators | Loads | Description |
|---|---|---|---|---|---|
| IEEE 30-Bus | 30 | 41 | 6 | 21 | Classic test case for verifying power flow algorithms on small networks. |

| IEEE 118-Bus | 118 | 186 | 19 | 99 | Widely used for PF studies involving voltage profiles and stability margins. |
| IEEE 300-Bus | 300 | 411 | 69 | 195 | Large-scale testbed for steady-state load flow benchmarking. |
| IEEE 1354-Bus | 1354 | 1991 | 260 | ~1100 | PEGASE system; designed for realistic, high-resolution PF analysis. |

These systems serve as the foundation for generating solved PF datasets under varied operating conditions, enabling a robust evaluation of the model's predictive accuracy and generalization capability across multiple grid scales.

### 3.2 Dataset Generation from IEEE Test Cases

Constructing a diverse and representative dataset for training and evaluating our GNN-based power-flow model, we generated multiple steady-state operating points for each IEEE test system. The generation process imitates realistic changes in system behavior stemming from load variation and generator dispatch. For each case (30, 118, 300, 1354-bus) we perturbed active and reactive demands at every load bus:

$$P_i^{new} = P_i^{base} \cdot (1 + \varepsilon_P), \qquad \varepsilon_P \sim \mathcal{U}(-0.4, 0.4) \tag{16}$$

$$Q_i^{new} = Q_i^{base} \cdot (1 + \varepsilon_Q), \qquad \varepsilon_Q \sim \mathcal{U}(-0.4, 0.4) \tag{17}$$

where $P_i^{base}$ and $Q_i^{base}$ are the nominal active and reactive power demands at bus $i$, and $\varepsilon_P$ and $\varepsilon_Q$ are random scaling factors simulating ±40% load variation. This setup captures both peak and off-peak scenarios, as well as seasonal and diurnal demand shifts. All power values are expressed in per unit (p.u.). [37]

For every simulation, generator voltage set-points were fixed at their nominal values, the slack bus absorbed any net power mismatch, and the power-flow solver was initialized from a flat start with all bus voltages set to 1.0 p.u. and phase angles to 0 rad, ensuring independence from prior solutions. Each perturbed case was then solved with the pandapower AC power-flow engine [38]; runs that failed to converge or produced infeasible voltage profiles were discarded to retain only physically realizable operating points. Repeating this process yielded 12,000 converged samples per test system, 5% of which included mild topology perturbations random N-1 line outages and, in meshed networks, and occasional transformer-tap adjustments so that the model would learn to predict bus voltages and angles not only across diverse load levels but also under credible reconfiguration scenarios.

Consequently, every operating point is exported as a graph object whose node-feature matrix **X** stores $[P_i, Q_i, V_i^{(init)}, \delta_i^{(init)}, PQ_i, PV_i, Slack_i]$ for each bus, whose edge list encodes the bidirectional connectivity of the physical network, and whose regression targets are the true voltage magnitudes $|V_i|$ and phase angles $\theta_i$ This representation retains the local electrical context of each bus while preserving the global topology required for message passing, enabling efficient processing within graph-neural architectures. After assembly, the full corpus 12,000 graphs per system, with 5% containing N-1 line-outage or tap-change perturbations was randomly partitioned into 70% training, 15% validation, and 15% test sets, ensuring that performance metrics reflect the model's ability to generalize to unseen operating conditions and topology variations.

### 3.3 Accuracy Metrics for Power-Flow Prediction

Model performance is evaluated using five standard regression metrics: mean-squared-error (MSE), root-mean-squared-error (RMSE), mean absolute error (MAE), normalized RMSE (NRMSE) and the coefficient of determination ($R^2$) [39].

MSE measures the average squared deviation between the predicted and true targets:

$$MSE = \frac{1}{N} \sum_{i=1}^{N} (\hat{y}_i - y_i)^2 \tag{18}$$

RMSE is obtained by taking the square-root of MSE, which restores the error to the original physical units:

$$RMSE = \sqrt{\frac{1}{N} \sum_{i=1}^{N} (\hat{y}_i - y_i)^2} \tag{19}$$

MAE computes the mean of the absolute deviations and is less sensitive to outliers:

$$MAE = \frac{1}{N} \sum_{i=1}^{N} |\hat{y}_i - y_i| \tag{20}$$

Since voltage magnitudes and angles have different scales across networks, we also report the range normalized RMSE (NRMSE):

$$NRMSE = \frac{\sqrt{\frac{1}{N} \sum_{i=1}^{N}(\hat{y}_i - y_i)^2}}{y_{max} - y_{min}} \qquad (21)$$

Finally, the coefficient of determination captures the fraction of variance explained by the surrogate:

$$R^2 = 1 - \frac{\sum_{i=1}^{N}(\hat{y}_i - y_i)^2}{\sum_{i=1}^{N}(y_i - \bar{y})^2} \qquad (22)$$

Lower values of MSE, RMSE, MAE and NRMSE together with higher $R^2$ (closer to 1) indicate superior predictive performance. A model is deemed better only when it achieves both smaller error metrics and a larger $R^2$ on the same test set.

### 3.5 GGNN-based Model: Training Convergence and Stability Analysis

In this section, we analyze the training dynamics of the best-performing model GGNN to shed light on convergence speed and stability on different system sizes. Figure 4 shows the training and validation loss curves for the GGNN on the 30-bus system and on the 1354-bus system. Each model was trained for a maximum of 800 epochs with early stopping based on validation loss.

In both cases, the GGNN converges rapidly to a very low loss. For the 30-bus network, the training loss (blue) and validation loss (orange) drop sharply within the first 20–50 epochs, reaching the $10^{-3}$ range, and thereafter continue to decrease gradually. The model achieved its best validation loss of approximately $4.45 \times 10^{-4}$ at epoch 737 (indicated by the early stopping criteria), with a final training loss of approximately $3.14 \times 10^{-4}$. This near-zero loss indicates an excellent fit to the 30-bus data without overfitting evidenced by the training and validation curves overlapping closely and flattening out together. On the 1354 bus system, a similar convergence pattern is observed: the loss plummets within the first 50 epochs to the $10^{-3}$ range, then slowly improves. The best validation loss reached on 1354 bus was about $9.61 \times 10^{-4}$ at epoch 790, with a final training loss of approximately $6.41 \times 10^{-4}$. We note that the 1354-bus model's loss is roughly double that of the 30-bus model's loss, which is expected since the larger system presents a more complex mapping and potentially a higher noise floor in the data. Importantly, the GGNN does not exhibit any instability or overfitting even on the large network, the validation loss closely tracks the training loss and remains low.

The convergence is smooth and monotonic in both cases, highlighting the effectiveness of the training procedure and the model's capacity. These training curves demonstrate that the GGNN can be efficiently trained on both small and very large grids, reaching a stable solution well before the maximum epoch with patience-based early stopping kicking in around the 740–790 epoch range. The model converges rapidly, achieving most of its loss reduction within the first few dozen epochs; this quick grasp of power-flow relationships makes it well-suited for practical deployment. On a standard 16-core CPU with a batch size of 16, per-epoch wall-clock time rises from a few seconds on the 30-bus case to well under a minute on the 1354-bus case; with early stopping, the entire training finishes in under an hour for the smallest network and in roughly seven hours for the largest, confirming that the GGNN can still be retrained overnight even when system size grows into the thousand-bus regime.

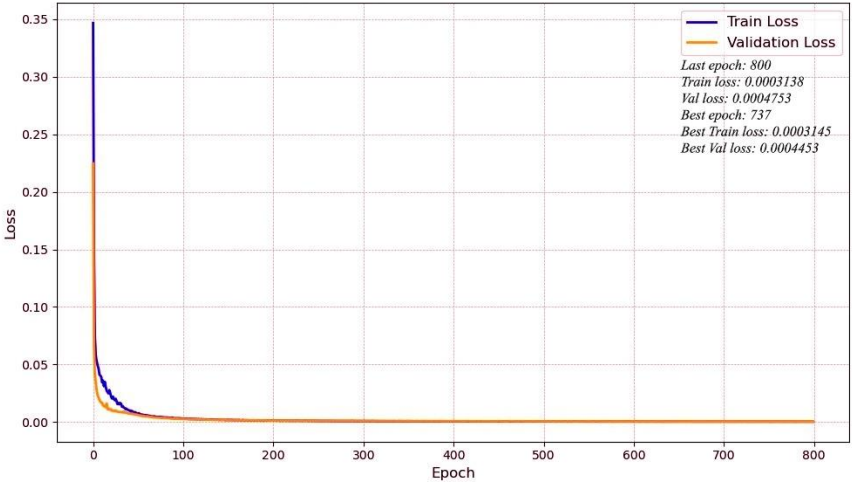

(a)

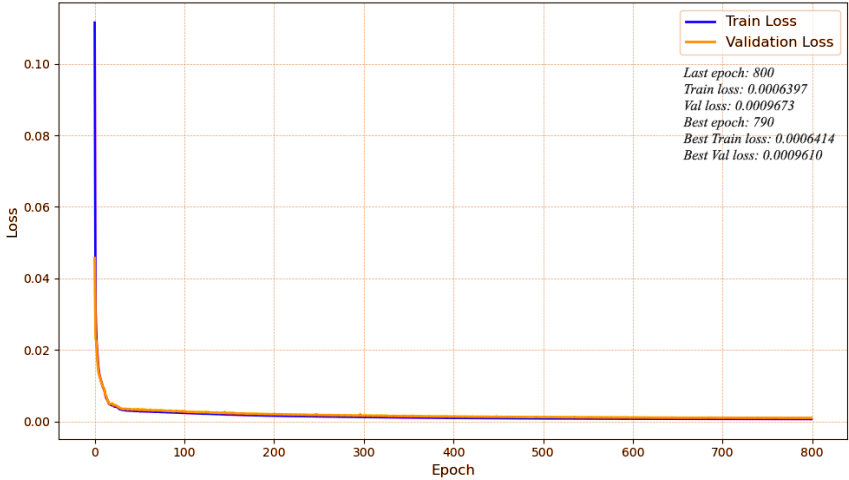

(b)

**Figure 4:** Training and validation loss curves for the GGNN on (a) the 30-bus system and (b) the 1354-bus system.

Finally, we examine the accuracy of the GGNN's predictions in terms of how closely they match the ground-truth power flow values. Figure 5 presents scatter plots comparing the predicted vs. actual bus voltage magnitudes and angles for the 30-bus system using the GGNN model. Each point in these plots corresponds to a bus voltage (magnitude or angle) in one of the test scenarios, and the color indicates the absolute prediction error. The plots reveal a high degree of alignment between predicted and actual values, with most points tightly clustering around the diagonal ideal prediction line. For voltage magnitudes, 1.64% of predictions fall outside acceptable operational bounds, while for voltage angles, the out-of-bound rate is only 0.03%. The points tightly cluster around the diagonal line dashed black line indicating perfect prediction, demonstrating the model's high accuracy. For voltage magnitudes, virtually all predictions fall very close to the actual values. The color of the points is predominantly deep purple (error < 0.002 p.u.), with only a few points in lighter shades; the error scale (right color bar) shows maximum errors on the order of only 0.008 p.u. This indicates that the GGNN's voltage magnitude predictions deviate by less than 0.2% at worst from the true values, and for most buses the error is a fraction of a percent. Similarly, for voltage phase angles, the predicted vs. actual plot shows points densely aligned along the diagonal, with an error color mostly below 0.002 radians.

The angle errors are all extremely small on the order of a few milliradians, with the color bar max at 0.004 rad equivalent to 0.23°). We observe no systematic bias in the errors – the scatter is symmetric around the line – meaning the model does not consistently over- or under-predict any region of the output space. The tight alignment of points in both subplots is reflected in a high $R^2$ value (for the 30-bus, $R^2$ about 0.956 for magnitudes and similarly high for angles), confirming that the GGNN captures most of the variance in the power flow outcomes. In summary, Figure 5 provides clear visual confirmation of the GGNN's predictive accuracy: the model's outputs match the actual power flow solutions with very small absolute errors, indicating that the GNN has learned a precise mapping from load conditions to the resulting grid state.

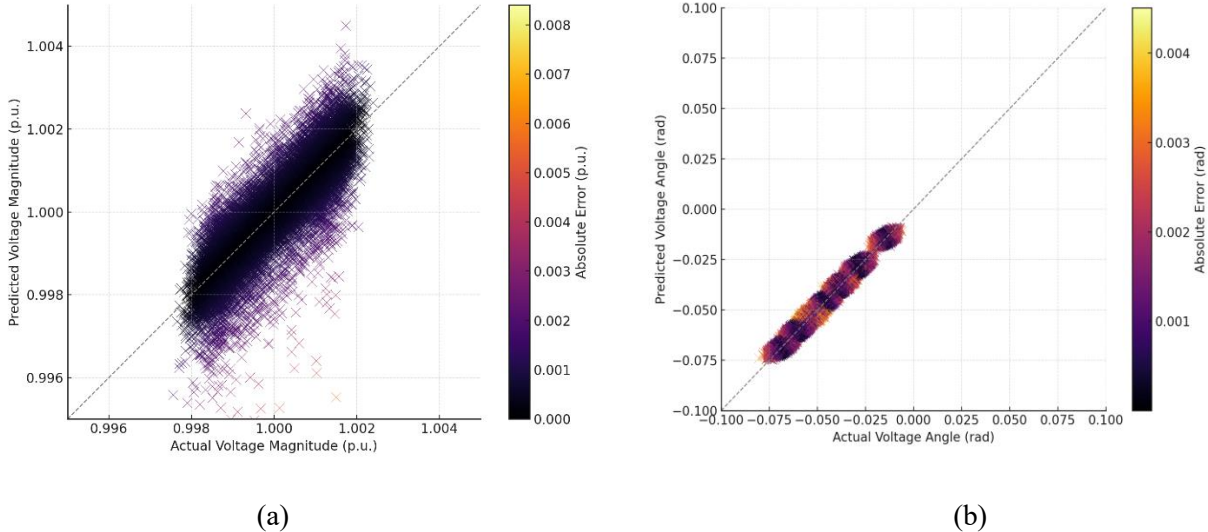

(a)            (b)

**Figure 5:** Predicted vs. actual values for the 30-bus system using GGNN, for (a) bus voltage magnitudes (per unit) and (b) bus voltage angles (radians).

## 3.4 GGNN-based Power Flow Estimation Across Different System Scales

As illustrated in Figure 6, a unified line chart benchmarks the performance of GGNN's across all four IEEE test systems using four key error-based metrics: MSE, RMSE, MAE, and NRMSE. On the 30-bus system, GGNN achieves excellent predictive accuracy, with an RMSE of 0.0223 and MAE of 0.0131. These low error values reflect the model's effectiveness in learning the mapping for smaller, well-conditioned grids and capturing localized power flow patterns with high precision. For the 118-bus system, GGNN continues to perform at a high level, with RMSE and MAE values of 0.0326 and 0.0180, respectively. The model maintains strong generalization as network complexity increases, demonstrating its ability to scale gracefully without degradation in estimation quality.

On the 300-bus system, GGNN delivers robust results, achieving an RMSE of 0.1775 and MAE of 0.0874. Despite the substantial increase in dimensionality and system interactions, the model effectively learns the complex spatial dependencies and nonlinearities of this large-scale grid, confirming its architectural scalability. In the 1354-bus system, GGNN maintains its strong performance, achieving some of its lowest error values: RMSE of 0.0309, MAE of 0.0160, and NRMSE of 0.0028. These results confirm that with sufficient data and careful training, GGNN generalizes reliably even on very large and realistic grid topologies, making it well-suited for practical deployment in real-world power systems. Overall, GGNN exhibits consistent, high-fidelity predictions across all system sizes. Its ability to generalize from small networks to complex, high-dimensional grids underscore its suitability for scalable, data-driven power flow estimation.

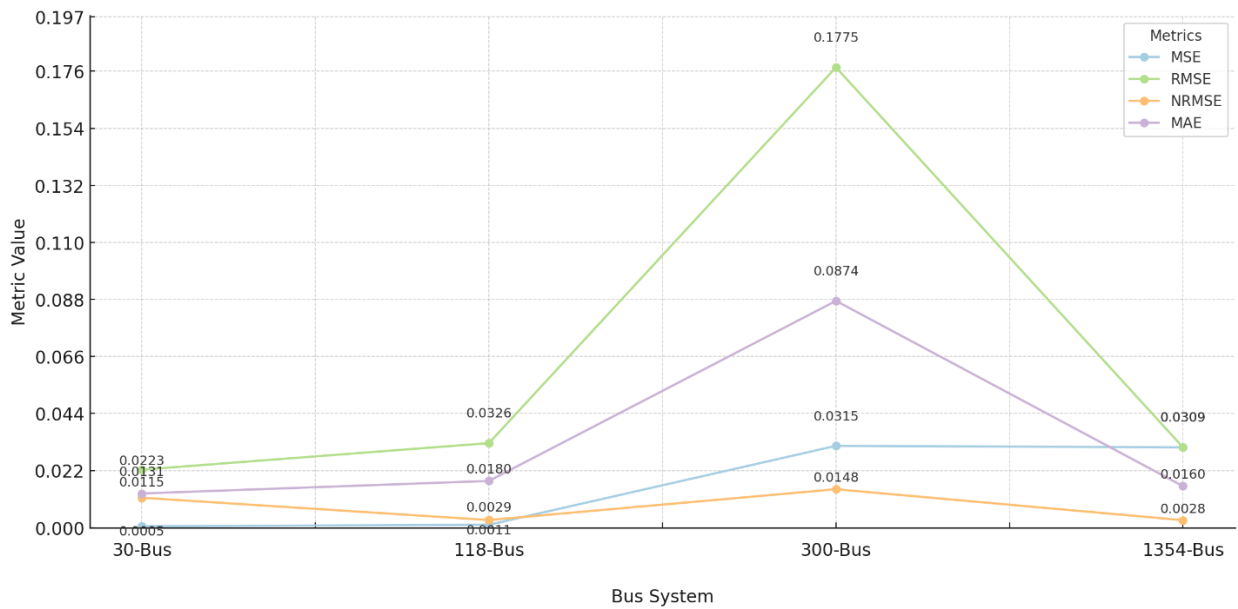

**Figure 6**: Performance of the Gated Graph Neural Network (GGNN) across IEEE test systems.

## 4. Detailed Comparisons Between the Proposed GGNN-based Framework and Existing Methods

### 4.1 Model Ranking Methodology

Each model was evaluated on four performance metrics MSE, RMSE, MAE, and R² on each test system. For each (model, system, metric) triplet, models were assigned rank scores, where a rank of 1 corresponds to the best performance and 14 to the worst. The ranks for all metrics and datasets were then summed and averaged to compute an overall Total Rank Score for each model. This approach ensures that the ranking system fairly rewards models that perform consistently well across different scales and evaluation criteria, while penalizing those that exhibit instability or metric-specific overfitting.

### 4.2 Ranking Results and Insights

Figure 7 shows the Average Total Rank for each model. The Gated Graph clearly emerges as the top-performing architecture with a total rank of 17.00, followed by TAGConv (21.50) and Transformer (24.75). These models consistently rank in the top 3 or 4 across most test systems and metrics, reflecting both accuracy and generalization strength. In contrast, models such as GCN, GAT, and MPNN rank significantly lower, with total scores above 50, indicating weaker and more inconsistent performance across scenarios. While some of these models may perform well under specific conditions or datasets, their lack of reliability across the full benchmark makes them less suitable for power system applications requiring robust performance.

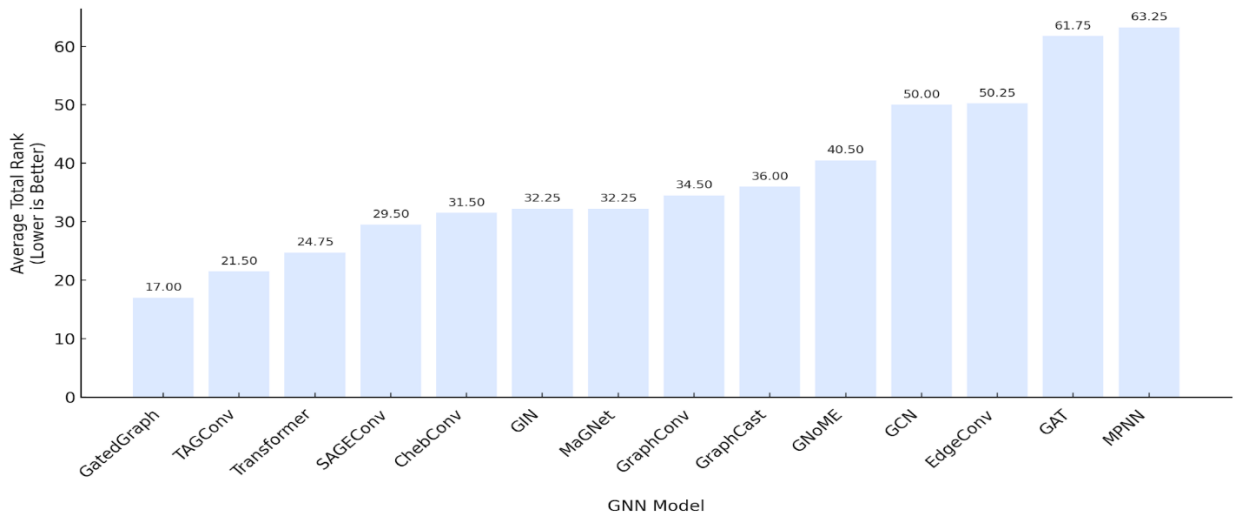

**Figure 7**: *Average total rank for each GNN model across all metrics and IEEE test systems. Lower values indicate better overall performance.*

## 4.3 Metric-Based Heatmap Analysis

To complement the rank-based comparison, Figure 8 presents a normalized heatmap of average metric scores for each model, aggregated across all datasets. Here, lower values (darker colors) represent better performance for MSE, RMSE, NRMSE, and MAE, while higher values indicate stronger $R^2$. GGNN shows dominant performance across all metrics, with particularly low normalized errors and strong $R^2$ scores. TAGConv and Transformer also demonstrate well-rounded profiles, although slightly less consistent than GGNN. Meanwhile, models like MPNN and GAT exhibit high normalized error scores, confirming their poor relative performance as reflected in the ranking analysis.

This dual view through rank aggregation and normalized metric heatmaps offers a holistic evaluation of model behavior. It confirms that a model's overall utility cannot be assessed by a single metric or dataset, but rather by its balanced performance across multiple criteria and grid complexities.

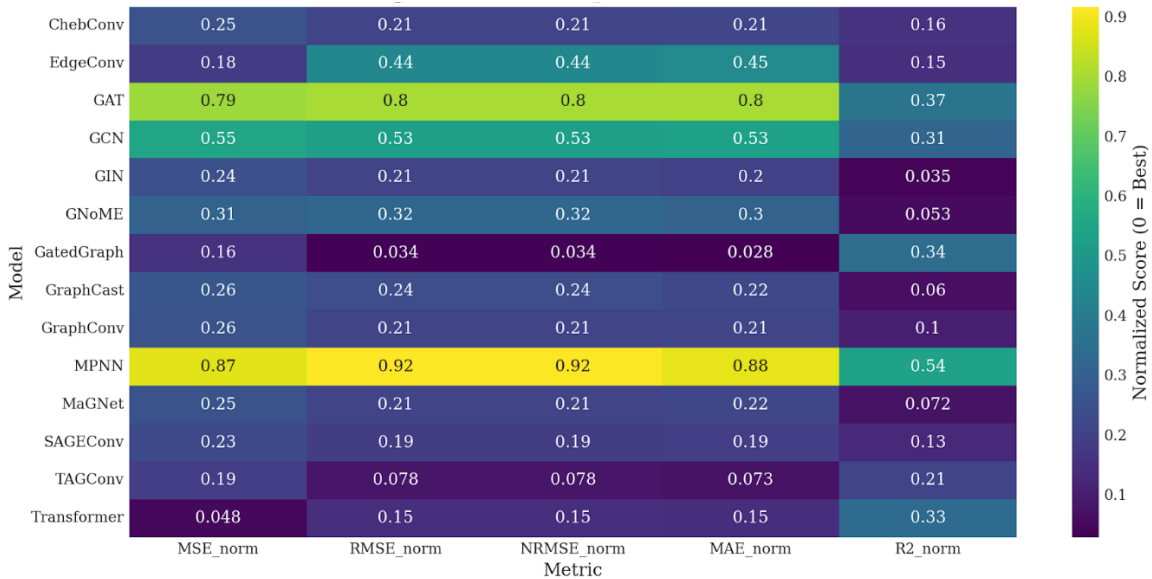

**Figure 8**: *Normalized performance heatmap across MSE, RMSE, NRMSE, MAE, and $R^2$ for each model.*

## 4.4 Per-Metric Performance Across Bus Systems

While the normalized heatmap gives an overall picture, it is instructive to inspect how each model's raw errors scale with the size of the power system. Figure 9 compares the MAE and RMSE obtained by all models on each of the four bus systems (lower values are better). In these bar charts, each group of four bars represents a single model's error on the 30, 118, 300, and 1354-bus test cases (red, green, orange, and blue bars respectively).

Several important trends are evident. First, most models exhibit increasing error as system size grows, the green and orange bars (118 and 300-bus) are generally taller than the red bars (30-bus) for a given model, reflecting the greater complexity of larger networks. However, by the 1354-bus system (blue bars), many models do not see a further spike in error, and in some cases the error even decreases slightly relative to the 300-bus case. This suggests that once a model successfully scales to a few hundred nodes, it can often handle a thousand-plus node system with comparable accuracy, likely due to the homogeneous nature of the task across scales (all in per unit). The superior performance of GGNN, TAGConv, and the Transformer is apparent across all systems – these models have the shortest bars (lowest RMSE and MAE) consistently. For instance, the GGNN's MAE remains under 0.02 p.u. even for 1354 buses, whereas most other models have significantly higher errors in the mid-size and large systems. TAGConv and Transformer show similarly low error profiles, indicating their effectiveness in capturing power flow relationships. In contrast, MPNN and GAT struggle on specific systems. The MPNN model has dramatic performance degradation on the 300-bus system (the orange bar for MPNN is by far the tallest among all models, with RMSE about 0.54 p.u. and MAE about 0.278 p.u.), revealing an inability to generalize to that network's complexity. GAT shows an unusually large error on the smallest 30-bus case (its red bar is high, corresponding to MAE about 0.046 p.u., much worse than other models on that system) suggesting that the graph attention mechanism did not suit the limited-data regime of the 30-bus scenario. Notably, by the 118-bus and larger systems, GAT's errors drop substantially (green/orange/blue bars for GAT are lower), implying it can learn effectively with more data, but its early performance was unstable. These per-system results reinforce the earlier ranking: GGNN, TAGConv, and Transformer not only achieve the lowest errors overall, but also remain robust as the network scale increases, whereas certain architectures like MPNN or GAT are prone to failing on at least one system (in MPNN's case, a severe failure at intermediate scale).

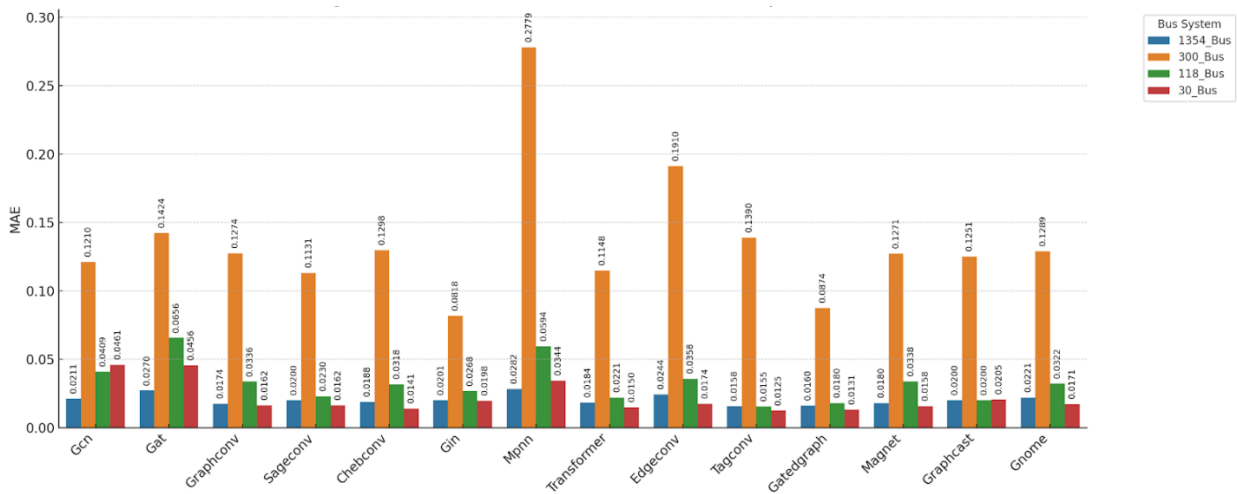

(a)

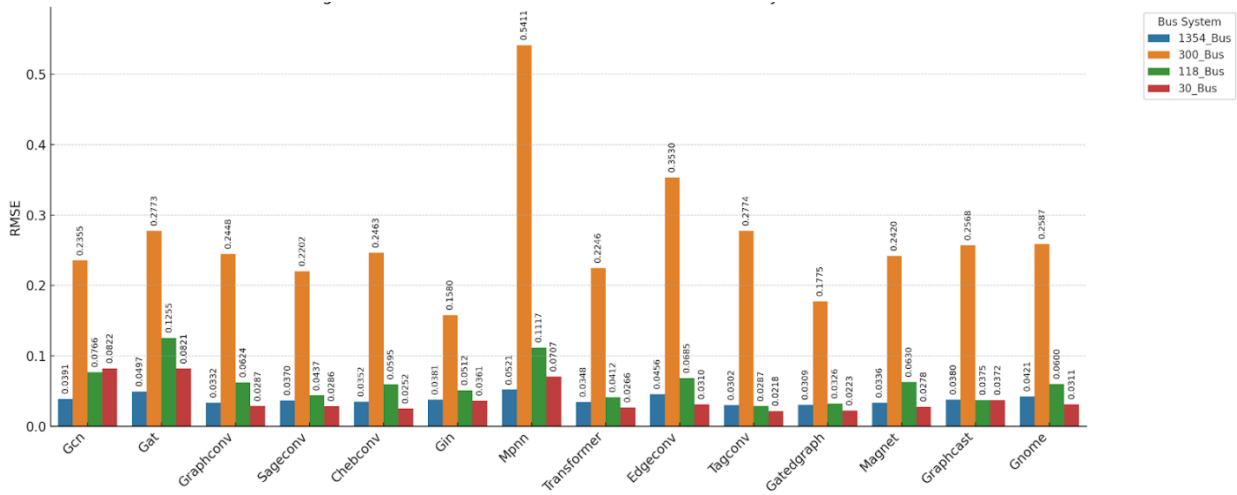

(b)

**Figure 9**: *Performance of all models in terms of (a) MAE and (b) RMSE on each bus system (30-bus in red, 118-bus in green, 300-bus in orange, 1354-bus in blue)*

## 4.5 Discussions and future work

Our evaluation across the IEEE 30-, 118-, 300-, and 1354-bus test systems shows that the GGNN consistently delivers the lowest prediction errors and $R^2$ scores above 0.99 on every metric examined (MSE, RMSE, MAE, NRMSE). TAGConv and a Transformer-based GNN also perform well, but they trail the GGNN slightly in aggregate ranking, while models such as MPNN and GAT reveal weaknesses on at least one system size. These results underscore the importance of assessing GNN architectures with a multi-metric, multi-system lens: an approach that seems reliable on one network or error measure may falter on another. The GGNN's rapid, stable convergence and its ability to scale from 30 to 1354 buses highlight its practicality for near-real-time grid-state estimation and other time-critical power-system applications.

Two key limitations should be acknowledged. First, the current surrogate relies solely on offline AC simulations; extending it to incorporate streaming PMU data or weather-driven renewable forecasts will require continual-learning mechanisms. Second, although our dataset covers single-line outages and load variability, it has yet to embrace more complex events e.g., generator trips or dynamic stability phenomena. Future work will therefore focus on embedding physics-informed constraints to ensure strict feasibility, coupling the GGNN with (security-constrained) optimal-power-flow pipelines, and deploying the model in hardware-in-the-loop testbeds to validate latency and robustness under realistic SCADA noise. Together, these efforts aim to transform GGNN-based surrogates into dependable, scalable components of next-generation energy-management systems.

## 5. Conclusion

This study presents a gated graph neural network (GGNN) surrogate model for AC power flow estimation and demonstrates its efficacy across diverse system conditions. The model is trained on IEEE benchmark power grids of various sizes and complexities, with each test case introducing random line contingencies and significant load variations to mimic realistic grid dynamics. Multiple training strategies are systematically explored to enhance model stability and generalization. The experimental results show that the GGNN surrogate yields highly accurate power flow predictions, consistently achieving lower RMSE and MAE and higher $R^2$ than conventional GNN-based surrogate models. These performance gains underscore the effectiveness of the gated architecture and training methodology, confirming the model's strong predictive fidelity and reliability under diverse operating scenarios.

Importantly, the GGNN model is designed to respect power system physics by embedding operational constraints within its architecture. By directly enforcing power balance and line loss relationships in the learning process, the model produces predictions that are physically consistent and robust. The model's accuracy remains strong even for increasingly large grid models, demonstrating the inherent scalability of graph neural networks for power system applications. In summary, the proposed GGNN surrogate delivers high-fidelity AC power flow estimates with superior accuracy, scalability, and efficiency, representing a significant advancement in data-driven power system modeling.